\documentclass{llncs}
\usepackage{makeidx}  
\usepackage[boxed]{algorithm2e}
\usepackage{psfig}
\usepackage{epsfig}
\usepackage{thfrm}
\usepackage{fancyhdr}
\usepackage[caption=false,font=footnotesize]{subfig}
\usepackage{verbatim}
\usepackage{float}
\usepackage{setspace}
\begin{document}
%
%
\pagestyle{headings}  
\addtocmark{Hamiltonian Mechanics} 
\mainmatter              

\title{Quantifying Privacy: A Novel Entropy-Based Measure of Disclosure Risk}
\titlerunning{A Novel Entropy-Based Measure of Disclosure Risk}  
%
\author{Mousa Alfalayleh \and Ljiljana Brankovic}

\institute{School of Electrical Engineering and Computer Science, The University of Newcastle, Callaghan, NSW 2308, Australia
\email{\{Mousa.Alfalayleh,Ljiljana.Brankovic\}@newcastle.edu.au}}

\maketitle

\begin{abstract}
It is well recognised that data mining and statistical analysis pose a serious treat to privacy. This is true for  financial, medical, criminal and marketing research. Numerous techniques have been proposed to protect privacy, including restriction and data modification. Recently proposed privacy models such as differential privacy and  k-anonymity received a lot of attention and  for the latter there are now several improvements of the original scheme, each removing some security shortcomings of the previous one. However, the challenge lies in evaluating and comparing privacy provided by various techniques. In this paper we propose a novel entropy based security measure that can be applied to any generalisation, restriction or data modification technique. We use our measure to empirically evaluate and compare a few popular methods, namely query restriction, sampling and noise addition.
\end{abstract}
\section{Introduction}

Proliferation of computer, network and communication technology and applications,
and in particular social networking and cloud computing had
a great impact on the way personal data is collected, stored and used. Data collected in one location (e.g., hospital) can now be stored
 remotely in a cloud and accessed from anywhere in the world.
These advances have undoubtedly changed the way we think about privacy~\cite{KingBrankovicGillard2012,Brankovic:Estivill-Castro:99,est-ca:bra:do} and
what once could have been regulated by legislative measures alone now requires
 a sophisticated suite of privacy enhancing technologies.
In this study we are concerned with a situation where confidential personal data is made
available  to a wide range of users who are authorised to perform data mining and statistical analysis, but not to access any
individual data.
There are
various \emph{Statistical Disclosure Control (SDC)} techniques that can be used to alleviate this
problem~\cite{Netherlands-Willenborg:ElementsSDC01,Nabil:ComparativeStudy89,Brankovic-Giggins:Ch12,Brankovic-Zahid:Ch11,FungWangChenYu2010} but, unfortunately, none
of them is able to solve it completely, due to  its
intrinsic contradictory nature. On one hand, one must keep the
risk of individual value disclosure as low as possible. On the
other hand, the utility (usefulness) of the data must remain
high. However, low risk implies low utility and high utility
implies high risk. A good SDC technique aims at finding a right
balance between the two. In order to achieve this balance, it is
crucial to adequately measure both utility and disclosure risk.
While measuring data utility has been well studied in the
literature ~\cite{bra:hor:mil:wri,bra:mil:sir02,bra,bra:sir,bra:mil:sir00,isl:bar:bra,Netherlands-Willenborg:ElementsSDC01}, measuring
disclosure risk is still considered as a difficult problem and has
been only partly solved.

\textbf{Contribution:} (1) In this paper we  propose a novel entropy
based measure of disclosure risk, which we refer to Confidential Attribute
Equivocation (CAE), and which is independent of the underlying SDC technique,
and thus can always be used. The main novelty and advantage of our technique over similar
ones is that it takes into account the candidate confidential values themselves,
rather than just their probabilities,  and is thus able to capture the risk of
approximate disclosure of confidential values, rather than the exact disclosure alone.
(2) We develop an efficient dynamic programming algorithm to evaluate  the CAE.
(3) We show how our technique can be applied to evaluate a few common SDC techniques,
including sampling, query restriction and noise addition.

This paper is organised as follows. In the next section we present
related work on disclosure risk measures and in Section 3 we
present a scenario that is not adequately covered by any of the
previous measures and  introduce a novel entropy-based
measure. In Section 4 we present a dynamic programming
algorithm for calculating the disclosure risk, in Section 5 we use
our measure to empirically evaluate a few existing SDC techniques, and in Section 6 we discuss
the experimental results and give some concluding remarks.

\section{Introduction to Statistical Disclosure Control and Related Work on Disclosure Risk Measures}

Privacy is an elusive concept, and many privacy models have been proposed,
with varying success. We next present a few of the most prominent models, including $k$-anonymity and differential privacy.

In a data set, some attributes may be considered public knowledge
and used to identify records. They are refereed to as
``identifying'' attributes or ``quasi-identifiers'' (QI).
A class of records where values of all QI attributes are the same is called
\emph{equivalence class}. To limit disclosure, Samarati and
Sweeney~\cite{Samarati:k-anonymity98}~proposed a so-called
\emph{k}-anonymity that requires each equivalence class to have no
less than $k$ records. The main problem with \emph{k}-anonymity arises
when all the records in an equivalence class share the same
confidential value, which allows intruder to disclose the
confidential value without actually re-identifying the record.
In order to alleviate this problem, Machanavajjhala et al proposed
 \emph{l}-diversity~\cite{MKGV07}, which in its simplest form  requires every QI to contain at
 least $l$ distinct values. While this model is indeed a great improvement over $k$-anonymity, it does not  consider how close these values are from each other,
 and thus leaves room for approximate disclosure of confidential values.
Li et al.~\cite{Ninghui:t-Closeness07}~introduced
\emph{t}-closeness, which considers the  distribution of
confidential attribute values in each equivalence class and the distribution of the confidential attribute values in
the whole dataset, and requires that the distance between these two distributions
does not exceed a given threshold \emph{t}. While it greatly reduces the disclosure risk, \emph{t}-closeness is overly restrictive and severely impacts the utility of data. In this context, our measure of disclosure risk can be seen as bridging a gap between the \emph{l}-diversity and the rigid \emph{t}-closeness.

Another prominent privacy model is differential privacy~\cite{Dwork2006}, which requires that the results to all queries allowed on the database do not change significantly if a single record is added or deleted from the database. While this is certainly an efficient model against table linkage attack, it is not design to
prevent attribute and record linkage \cite{FungWangChenYu2010} and in practice may be inferior to k-anonymity \cite{sra:saf-nai:den:ask}.


Each one of the above models can be implemented using different SDC techniques, which can be classified as modification
techniques and query restriction
techniques~\cite{Brankovic-Zahid:Ch11,Brankovic-Giggins:Ch12,Netherlands-Willenborg:ElementsSDC01,Denning:Cryptography82}.
\emph{Modification techniques} involve some kind of alternation of the
original data set before it is released to statistical users. This
includes noise addition, data swapping, aggregation, suppression
and sampling~\cite{Brankovic-Giggins:Ch12,Brankovic-Zahid:Ch11,isl:bra,bra:lop:mil:seb,isl:bra:05}. The common denominator
of all modification techniques is that the modified dataset is
released to users who are free to perform any query on it, but the
answers they get are only approximate and not exact. On the other hand,
\emph{query restriction techniques} do not release
database to a user but rather provide a query access. The SDC
system decides whether or not to answer the query but if the query
is answered, the answer will always be exact and not approximate
as with modification techniques~\cite{Denning:Cryptography82,Brankovic-Giggins:Ch12}.

In this study we are not concerned with SDC techniques or  privacy models as such but rather with
measuring disclosure risk. In the literature, disclosure risk measures are
classified as
measures for record re-identification or confidential value
disclosure~\cite{Duncan:dissemination86,Lambert:Measures93,Brankovic-Giggins:Ch12}.
The latter focuses on measuring the risk of compromising a
confidential value of a particular individual, while the former
focuses on measuring the risk of inferring an individual's
identity. In either case the disclosure risk measures may
be applied to the database as a whole, or  to
individual records.

Several methods have been proposed to estimate the disclosure risk
in sampling and they fall under the category of record
identification. Winkler~\cite{Winkler:Masking04}~refers to these
methods as Sample-Unique-Population-Unique (SUPU) methods as
disclosure risk estimation requires assessing the uniqueness of
records in the released sample and in the population. Skinner and
Eliot~\cite{Skinner-Elliot:Measure02}~introduced a new disclosure
risk measure for microdata which falls under SUPU
methods~\cite{Winkler:Masking04}. Their measure is based on the
probability $\theta=Pr(\mbox{correct match}|\mbox{unique match})$
that a microdata record and a population unit are correctly
matched.
Additionally, they introduced a
simple variance estimator and  claimed that their measure is
able to evaluate the different ways of releasing microdata from a
sample survey. Truta et
el.~\cite{Truta-Fotouhi:Disclosure04}~introduced other SUPU
measures and termed them minimal, maximal, and weighted
disclosure risk measures. The minimal disclosure risk measure is
the percentage of records in a population that can be correctly
re-identified by an intruder. All these records must be population
unique. The maximum disclosure risk measure takes into account
records that are not population unique while the weighted
disclosure risk measure assigns more weights to unique records
over other records. Their measures are not linked to a certain
individual but compute the overall disclosure risk for the
database. These measures can only be applied to limited SDC
methods such as sampling and microaggregation and it is considered
hard to choose the disclosure risk weight
matrix~\cite{Truta-Fotouhi:Disclosure04}. However, assigning
weights enables a data owner to setup different levels of
confidentiality. These measures are useful in deciding the order
of applying more than one SDC method on the initial data.

Trottini and Fienberg~\cite{Trottini-Fienberg:Modelling02}~proposed
a simple Bayesian model for capturing user uncertainty after
releasing the data by an agency. They distinguish between the
legitimate user (researcher) uncertainty and the malicious user
(intruder) uncertainty. This distinction is used as the basis of
defining appropriate disclosure risk measure. The proposed measure
is an arbitrary decreasing function of the user's uncertainty
about a confidential attribute value.

Spruill~\cite{Spruill:Measures82}
 measured confidentiality as a percentage of records in the
released data where a link with the original
 data can not be made. In
order to decide if there is such a link, for each released record,
we add up either the square or the absolute value of the difference between the released value and the true
value for all common numerical attributes. A link is said to be made if a
released record was derived from the true record that has the
minimum sum of differences. Spruill's early work gave rise to \emph{record linkage}, much studied in recent years~\cite{FungWangChenYu2010}.

There are some recent proposals that use information-theoretic approach
to measure privacy and utility of various SDC techniques
\cite{Oganian-Ferrer:Posteriori03,san:raj:poo}; however, none of them
measures the ``approximate" compromise, as we explain in the next section.

\section{A Novel Entropy-Based Measure}
Out of all disclosure risk measures, the closest to our proposal is a measure proposed by Onganian and Domingo-Ferrer~\cite{Oganian-Ferrer:Posteriori03} that evaluates the security of releasing
tabular data. The measure is equal to the reciprocal of
conditional entropy given the knowledge of an intruder:
\begin{equation}\label{1}
    DR(X)=\frac{1}{H(X|Y=y)}=\frac{1}{(-\sum_{x}p(x|y)\log_{2}p(x|y))}
\end{equation}
where \emph{X} represents a confidential attribute for a given
record and \emph{Y} represents intruder's knowledge. The
disclosure risk is inversely proportional to the uncertainty about
the confidential attribute given intruder's knowledge. The measure
performs \emph{a posteriori}, that is, after applying one of SDC
methods to the tables. It is a complement to \emph{a priori}
measures such as some currently used sensitivity rules including
(\emph{n,k})-dominance and \emph{pq}-rule, which help a data owner
in deciding whether to release the data or not. The main strength
of the above a posteriori measure is its generality: it is
applicable to various SDC methods such as Cell Suppression,
Rounding, and Table Redesign. In order to evaluate this disclosure
risk, one has to find a set of the possible confidential attribute
values and their probabilities given the condition $Y=y$. A down
side of this measure is that it does not capture accurately the
knowledge that an intruder has about a confidential attribute, as
it does not give careful consideration to the attribute values but
only the probabilities with which the values occur. Our proposed
method considers the attribute values in addition to their
probabilities.

Before we proceed to describe our measure in more detail and compare some of
the SDC techniques in the experimental section, we need to
introduce the concept of ``database compromise''. We say that a
database is \emph{compromised} if a \emph{sensitive statistic} is
disclosed ~\cite{Denning:Cryptography82}. There are several
distinct types of compromise, depending upon what is considered to
be sensitive. For example, if only exact individual values are
considered sensitive, we have the so-called \emph{exact
compromise}.
 \emph{Approximate compromise} occurs when a user is
able to infer that a confidential individual value $X$ lies within
a
range~$[X_{0}-\frac{\huge{\varepsilon}}{2},X_{0}+\frac{\huge{\varepsilon}}{2}]$~for
some predefined value of $\huge{\varepsilon}$. Approximate
compromise will prove crucial for the definition of our new
security measure.


We consider a scenario where an intruder is trying to unlawfully
disclose confidential information from a database. She uses
all the available information she can get from the database, as well as any external knowledge she may have. At the
end of her analysis, the intruder is able to reduce the possibilities
and limit her suspicion to certain data values. Shannon's entropy can measure the
intruder's uncertainty, but does not
take into consideration how far or close these values are from
each other.
%

\begin{table}[h]
\begin{center}
\footnotesize{
\caption{Two Examples} \label{table:QueryExample}
\begin{tabular}{| l | l |l|}
\hline & \textbf{Example 1} & \textbf{Example 2}\\
\hline Q1: & SELECT~MAX(Salary) & SELECT~MAX(Salary) \\
& FROM~AcademicStaff & FROM~AcademicStaff\\
& WHERE~Sex = "F" & WHERE~Age = $37$\\
\hline A1: & Maximum salary = $107,000$ & Maximum salary = $80,000$ \\
\hline Q2: & SELECT~AVG(Salary) & SELECT~AVG(Salary) \\
& FROM~AcademicStaff & ~FROM~AcademicStaff\\
& WHERE~Sex = "F" & ~WHERE~Age = $37$\\
\hline A2: & Average salary = $78,500$ & Average salary = $78,500$\\
\hline
\end{tabular}}
 \end{center}
\end{table}
The first example in Table~\ref{table:QueryExample}~ shows the
queries submitted by the intruder and the database response to
them. Assuming that there are only two female academics, Layla and
Angela, the intruder learns that Layla's salary is one of the two
values: It is either $107K$ or $50K$.

In the second example we assume that there are only two academics aged $37$,
Qays and Tony. Then the intruder knows that Qays's salary
is either $80K$ or $77K$. If we  use Shannon's entropy to evaluate the intruder's uncertainty  in examples $1$ and $2$, we get
 the same result, 1 bit in each case. However, we argue that the
intruder learns more in example $2$, as he can pretty accurately
estimate the salary to be $78.5K \pm 1.5K$. This highlights the
need for more accurate measure than Shannon's entropy, which would
be able to capture such differences.

Recall that a compromise can be exact or approximate.  Shannon's entropy
can be considered a satisfactory measure for the disclosure risk
that is related to the exact compromise. However, in the
approximate compromise, we argue that Shannon's entropy does not
express precisely the intruder's knowledge about a particular
confidential value.

We introduce a notion of privacy for the
so-called \emph{approximate compromise range}
($\huge{\varepsilon}$). In the approximate compromise an intruder
learns that the confidential value $X$ lies within a range
$[X_{0}-\frac{\huge{\varepsilon}}{2},X_{0}+\frac{\huge{\varepsilon}}{2}]$.
For the two example  we  have $X \in
[X_{0}-28.5K,~X_{0}+28.5K]$ for Layla and $X  \in
[X_{0}-1.5K,~X_{0}+1.5K]$ for Qay, where in both cases $X_{0}=78.5K$.
Obviously, the intruder knows more about Qay's than Layla's salary,
as in the former the approximate compromise range is $3K$,
while in the latter it is $57K$.

\begin{figure}[]
       \begin{center}
            \includegraphics [scale=0.35]{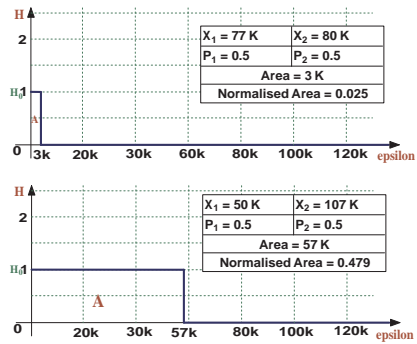}
            \caption{Our Security Measure: for (77K, 80K) and (50K, 107K).} \label{fig:OurMeasure77-80}
       \end{center}
   \end{figure}

To capture the information about the range $\huge{\varepsilon}$,
we  use Shannon's entropy $H$ as a function of $\huge{\varepsilon}$.
The graphs in
Figure~\ref{fig:OurMeasure77-80}~correspond to the intruder's
uncertainty $H(\huge{\varepsilon})$ in the above examples. We notice that
the entropy $H(0)$ is the same in both cases, that is, the disclosure risks are the same for exact compromise.
However, the area under $H(\huge{\varepsilon})$ is
much larger for Layla  implying that this case is more resistant against approximate compromise.

In general, we can then evaluate
intruder's uncertainty for any given $\huge{\varepsilon}$. In particular, we use $H_{0} = H(0)$ to
denote intruder's uncertainty in the case of exact compromise, that is, approximate compromise
range of ``$0$'' and we call it \emph{initial entropy}. Additionally, in what follows we
 examine the area ($A$) determined as an integral: $A=\int_{0}^{\huge{\varepsilon}_{max}}
 H(\huge{\varepsilon})$, where ``$\huge{\varepsilon}_{max}$'' is the value of $\huge{\varepsilon}$
for which entropy $H(\huge{\varepsilon})$ drops to zero. Formally, $H(\huge{\varepsilon}_{max}=0)$
and $H(\huge{\varepsilon}>0)$ for all $\huge{\varepsilon}<\huge{\varepsilon}_{max}$

We next  explain  how $H(\huge{\varepsilon})$
is calculated in general. We introduce a ``window'' of length
$\huge{\varepsilon}$. When a window ``covers'' two or more values,
then they are replaced with a single value whose probability is
equal to the sum of probabilities of all the values covered by the
window. In general, there will be more than one way to cover the values with
windows of length $\huge{\varepsilon}$ and we need to select the way that minimises the entropy $H(\huge{\varepsilon})$.
We next formally define the problem of calculating $H(\huge{\varepsilon})$.

\textbf{Calculating $H(\huge{\varepsilon})$}\\
\textbf{ Input}:\\
* Sorted values  $x_1,x_2,\dots,x_n$, $x_i > x_j$ for any $i > j$\\
* Their corresponding probabilities: $p_1,p_2,\dots,p_n$\\
* Parameter $\huge{\varepsilon} \leq x_n-x_1$  \\
\textbf{Output}:\\
* Collection $C$ of subsets $(x_1,...x_{y_1}), (x_{y_1+1},...,x_{y_2}), \dots, (x_{y_m+1},...,x_n)$,  such that (1) $x_{y_1}-x_1 \leq \huge{\varepsilon}$,  (2) $x_{y_i}-x_{y_{i-1}+1} \leq \huge{\varepsilon}$, $2 \leq i \leq m$, (3) $x_n-x_{y_m+1}  \leq \huge{\varepsilon}$\\
* Corresponding probabilities $q_1=p_1+...+p_{y_1}, ..., q_{m+1}=p_{y_m+1}+...+p_n$\\
* $H(\huge{\varepsilon})$ calculated over probabilities $q$, such that $H(\huge{\varepsilon})$ is maximised over all collections $C$   satisfying conditions above.

Computing the minimum entropy $H(\huge{\varepsilon})$ as a function of
$\huge{\varepsilon}$ is not straightforward and in Appendix
we provide an example to illustrate the entropy calculation.
In the next section we introduce a dynamic programming algorithm to find
the optimal entropy and hence calculate the area ($A$) that
together with the initial entropy ($H_{0}$) represents our
disclosure risk.

\section{A Dynamic Programming Algorithm to Compute~$H(\huge{\varepsilon})$}

We are given as input a set of values $x_{i}$ in increasing order
($x_{1}<x_{2}<x_{3}<\cdots<x_{n}$) where each $x_{i}$ has a given
probability $p_{i}$ where $p_{i}~\geq~0$ and and $\Sigma p_{i}=1$.
In order to produce our security measure, we need to calculate
minimum $H(\huge{\varepsilon})$ for each $\huge{\varepsilon}$. We
break the problem into stages (rows) and states (columns). Each
row in the table corresponds to a stage or $\huge{\varepsilon}$. Column ``$i$''
in the table corresponds to  the subproblem containing values
$x_{1},~x_{2},\cdots,~x_{i}$. For a given row (stage) in the
table, each cell in this row is viewed as a subproblem
~$H(\huge{\varepsilon},i)$~of the original
problem~$H(\huge{\varepsilon})$. For a given stage and state,
~$H(\huge{\varepsilon},i)$~ is computed by the following
recurrence:
$$H(\huge{\varepsilon},i)=min[(H(\huge{\varepsilon},i-1)+a_{i}),(H(\huge{\varepsilon},i-2)+a_{i-1}),...,(H(\huge{\varepsilon},j-1)+a_{j})]$$

\noindent
where $$a_{j} = (\sum_{k=j}^{i} p_{k})\cdot log (\frac{1}{(\sum_{k=j}^{i} p_{k})})$$
\begin{center}
$X_{i}-X_{j} \leq  \huge{\varepsilon}$, $X_{i}-X_{j-1}>\huge{\varepsilon}$ for  $i,j \in [1,n]$ and  $i \geq j$
\end{center}
 $$H(0,\huge{\varepsilon})= 0$$
 $$H(1,\huge{\varepsilon})= p_{1}\cdot log (\frac{1}{p_{1}})$$



\begin{algorithm}[H] \footnotesize
  \SetLine
  \KwIn{$x[~]$: a set of integer values in ascending order\;
  ~~~~~~~~~~~$p[~]$: a set of probabilities corresponding to the above integer values.}
  \KwOut{$H(\huge{\varepsilon})$}

  $H_{0} \leftarrow  \sum_{i=1}^{n} p(x_{i})\cdot log (\frac{1}{p(x_{i})})$\;
  \ForEach{$\huge{\varepsilon}$}{
    $H(\huge{\varepsilon},0) \leftarrow 0$\;
    $H(\huge{\varepsilon},1) \leftarrow p_{1} \cdot \log(\frac{1}{p_{1}})$\;
    \For{$i\leftarrow 2$ \KwTo $n$}{
        $j \leftarrow i$\;
        $p_{partial} \leftarrow 0$\;
        $H(\huge{\varepsilon},i) \leftarrow H(\huge{\varepsilon-1},i) $\;
        \While{$(x_{i}-x_{j} \leq \huge{\varepsilon})~and~(j \neq 0)$}{
            $p_{partial} \leftarrow p_{partial}+p_{j}$\;
            $H_{temp} \leftarrow p_{partial} \cdot \log(\frac{1}{p_{partial}}) + H(\huge{\varepsilon},j-1)$\;
            \If{$H_{temp} < H(\huge{\varepsilon},i)$}{
                $H(\huge{\varepsilon},i) \leftarrow H_{temp}$\;
            }
            $j \leftarrow j-1$\;
        }
    }
    $H(\huge{\varepsilon}) \leftarrow H(\huge{\varepsilon},n)$\;
    \textbf{Display:} $H(\huge{\varepsilon})$
  }

  \caption{A dynamic programming algorithm to compute~$H(\huge{\varepsilon})$}\label{algorithm:DPALGO1}
\end{algorithm}
\section{The Experiments: Description and Implementation}
In this section we apply our proposed security measure to a few
common Statistical Disclosure Control (SDC) techniques. In all
instances we assume that the intruder has \emph{supplementary
knowledge} (SK) about an individual whose corresponding record is
stored in the original dataset, which  can be as limited as one
attribute or can be as extensive as all attributes except the
confidential one.


The comparative study is performed on three different data sets.
Here we present results on PUMS dataset, whose decription can be found in the Appendix.

\textbf{Sampling.} In this experiment, we release a subset of records (sample) from
the original microdata file (population). We use a simple random
sample without replacement where each record in the original is
equally likely to be included in the produced sample and
duplicates are not allowed. The produced sample has a constant
size specified as a percentage of the original dataset total size
and referred to as ``sampling size" (or ``sampling factor''). In deciding on the
structure of the sampling experiment, we follow work by Truta et
el.~\cite{Truta-Fotouhi:Disclosure04}~on disclosure risk measures
for sampling, where we compute the overall disclosure risk for the
database, rather than for a certain
individual.

%
%

\begin{figure}[H]
\footnotesize{
    \raggedleft
        \subfloat[Sampling: $H_{0}$ vs SK ]{%
                    \label{fig:Sampling_USA_General_H0_vs_AttrNumber}%
                    \includegraphics[width=0.33\textwidth]{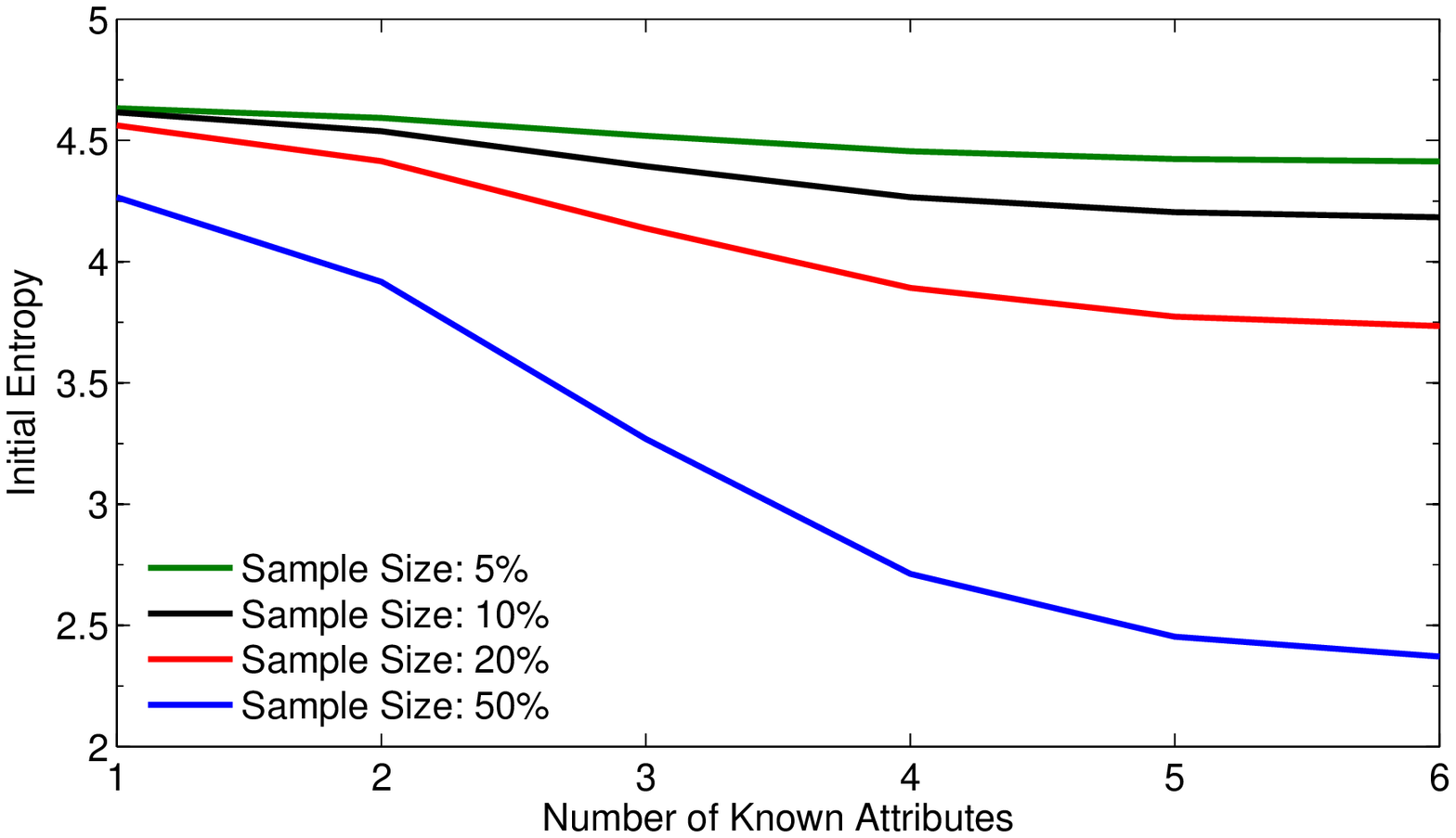}  }%
        \subfloat[Sampling: Area vs SK ]{%
                    \label{fig:Sampling_USA_General_Area_vs_AttrNumber}%
                    \includegraphics[width=0.33\textwidth]{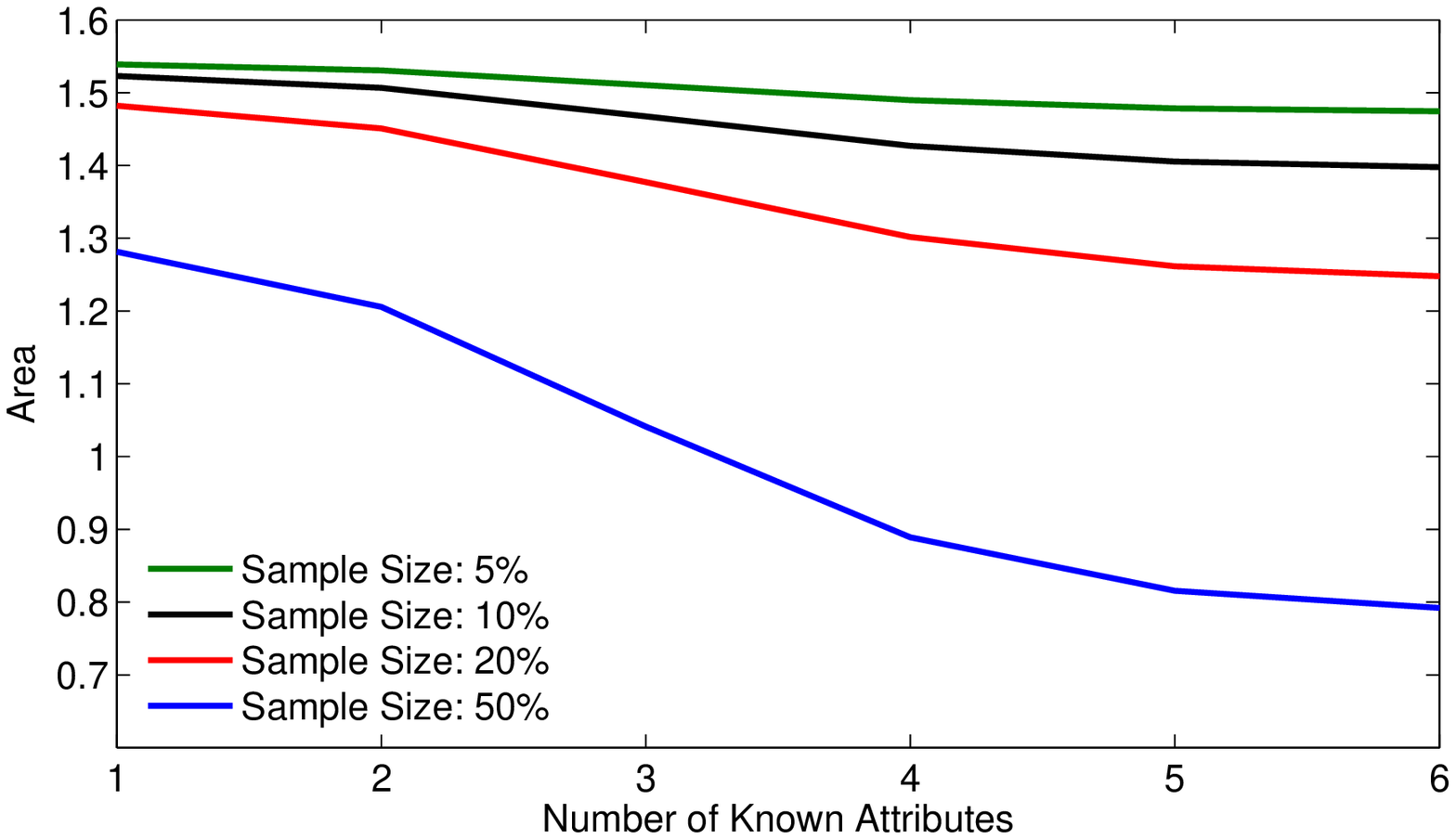} }
        \subfloat[QueryRestr.: $H_{0}$ vs SK]{%
                    \label{fig:QR_USA_General_H0_vs_AttrNumber}%
                    \includegraphics[width=0.33\textwidth]{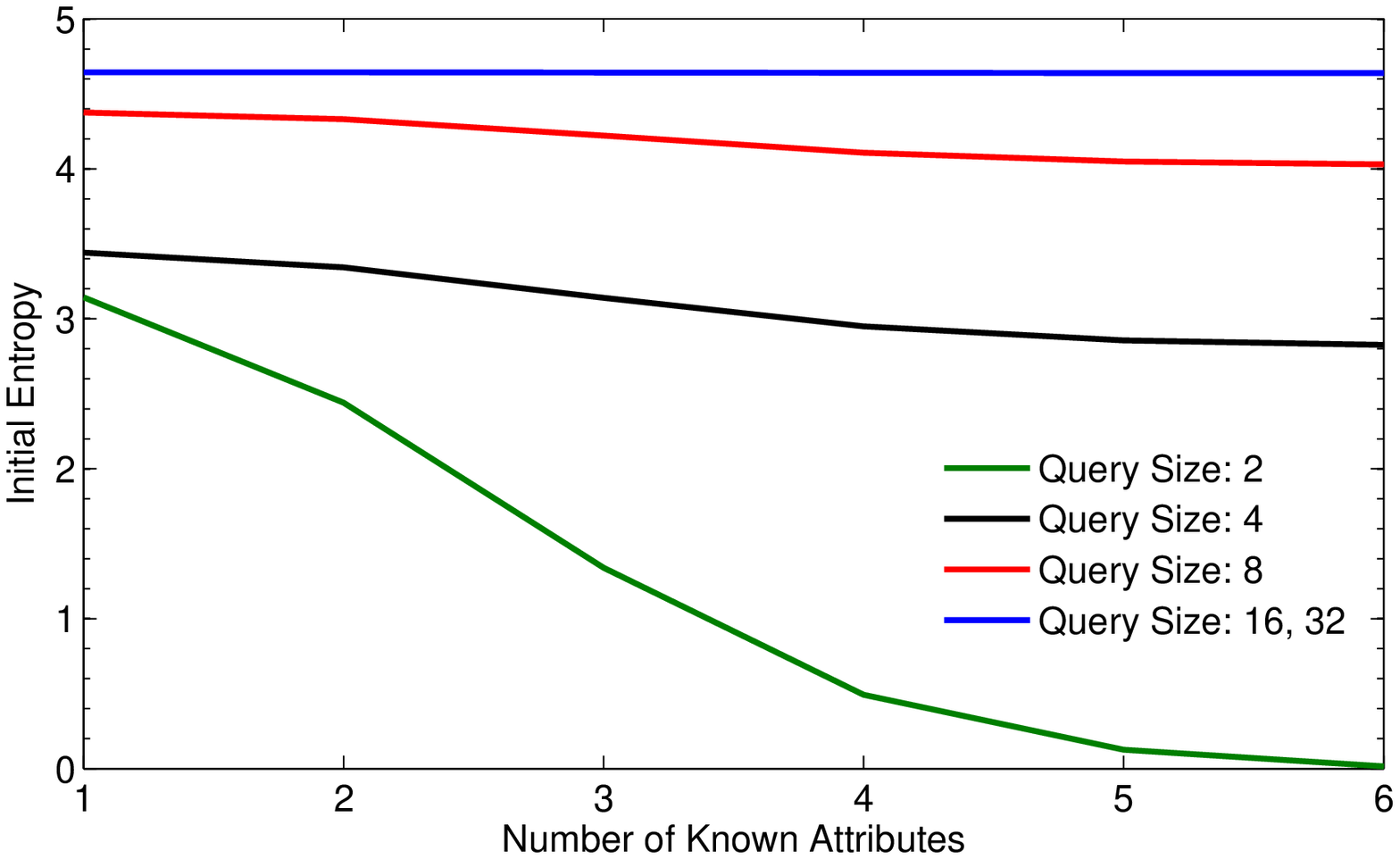}  }\\
        \subfloat[QueryRestr.:Area vs SK ]{%
                    \label{fig:QR_USA_General_Area_vs_AttrNumber}%
                    \includegraphics[width=0.33\textwidth]{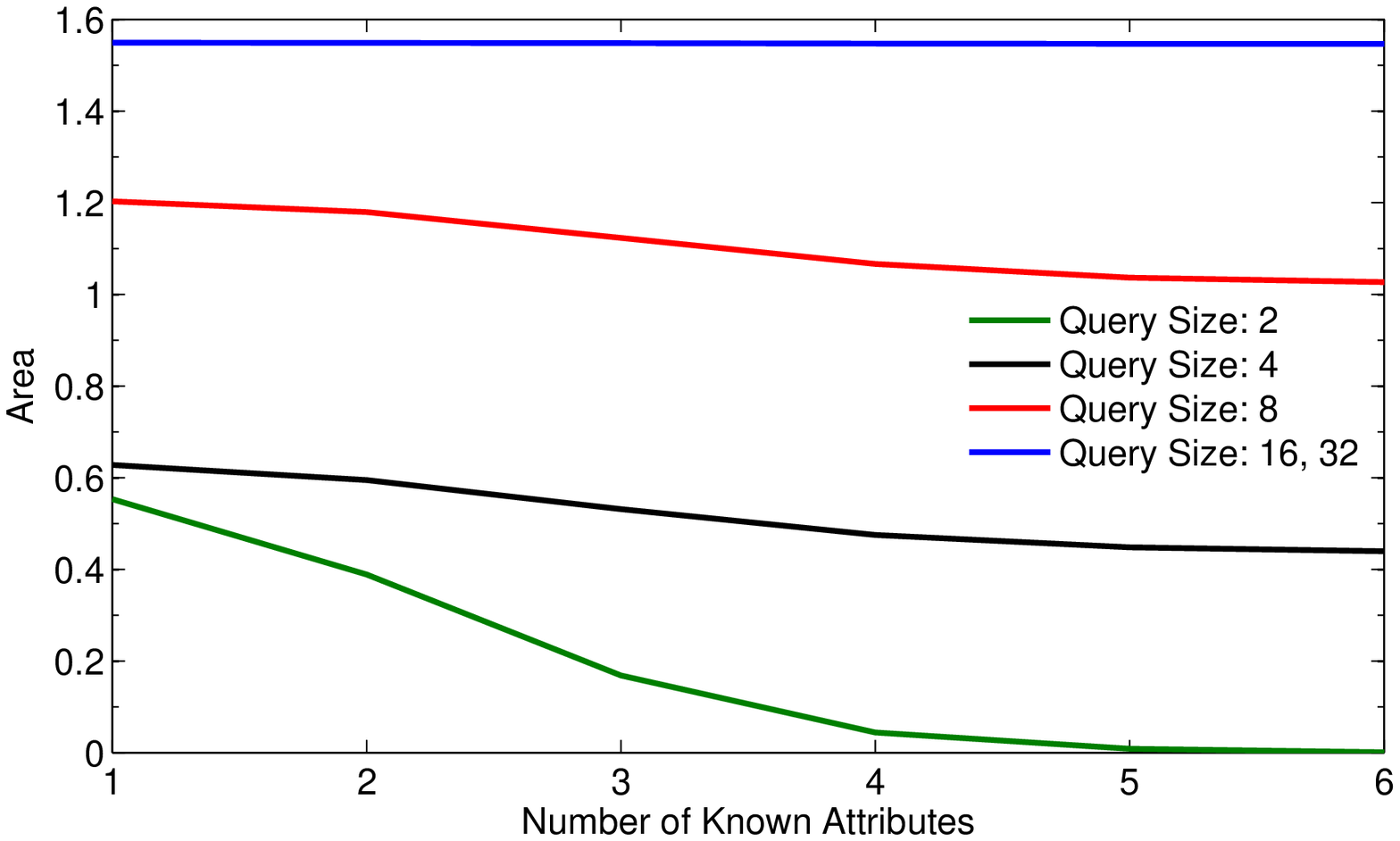}
                    }
        \subfloat[NoiseAdd.: $H_{0}$ vs SK ]{%
                    \label{fig:NA_USA_General_H0_vs_AttrNumber}%
                    \includegraphics[width=0.33\textwidth]{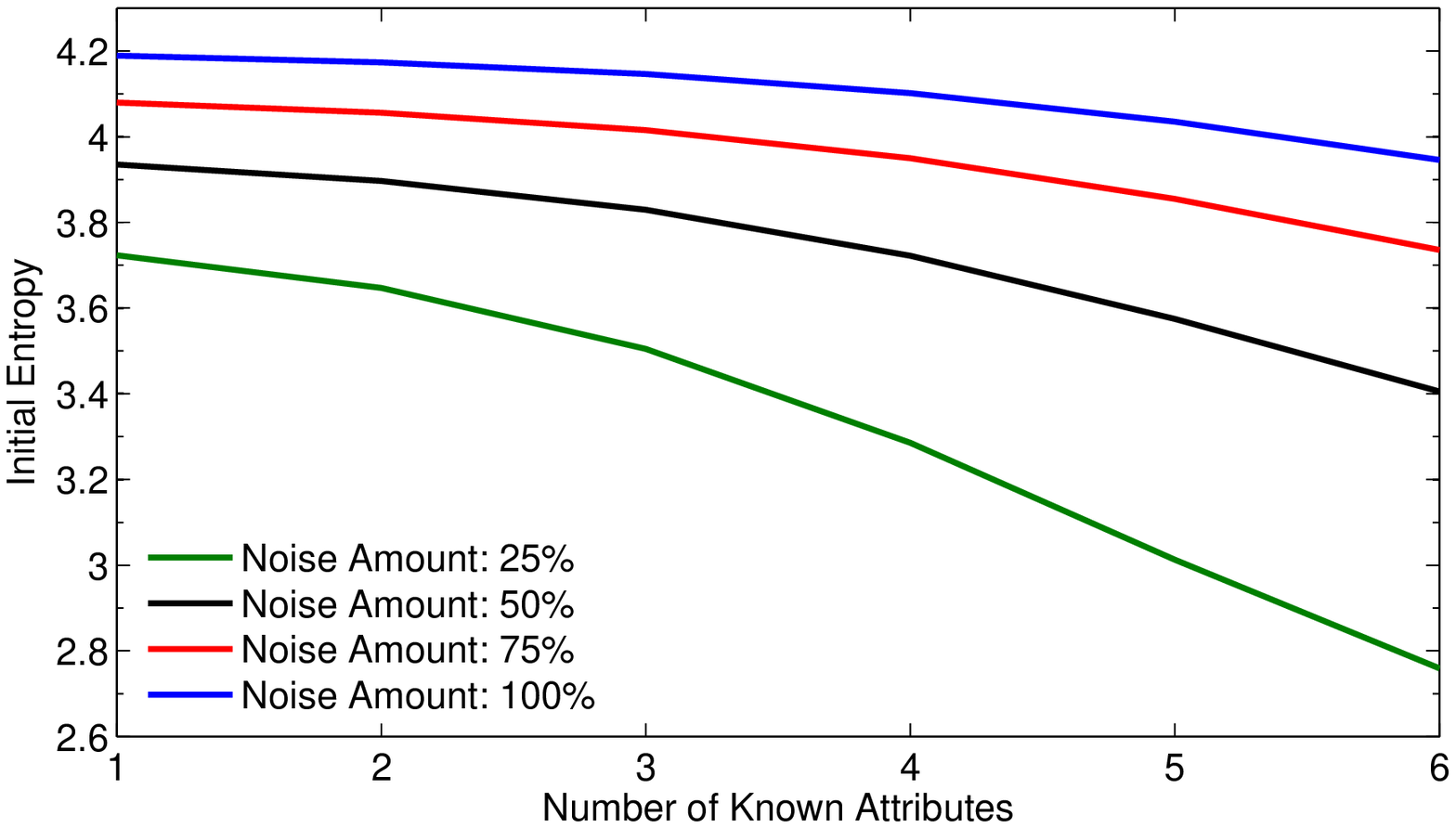}  }%
        \subfloat[NoiseAdd.: Area vs SK ]{%
                    \label{fig:NA_USA_General_Area_vs_AttrNumber}%
                    \includegraphics[width=0.33\textwidth]{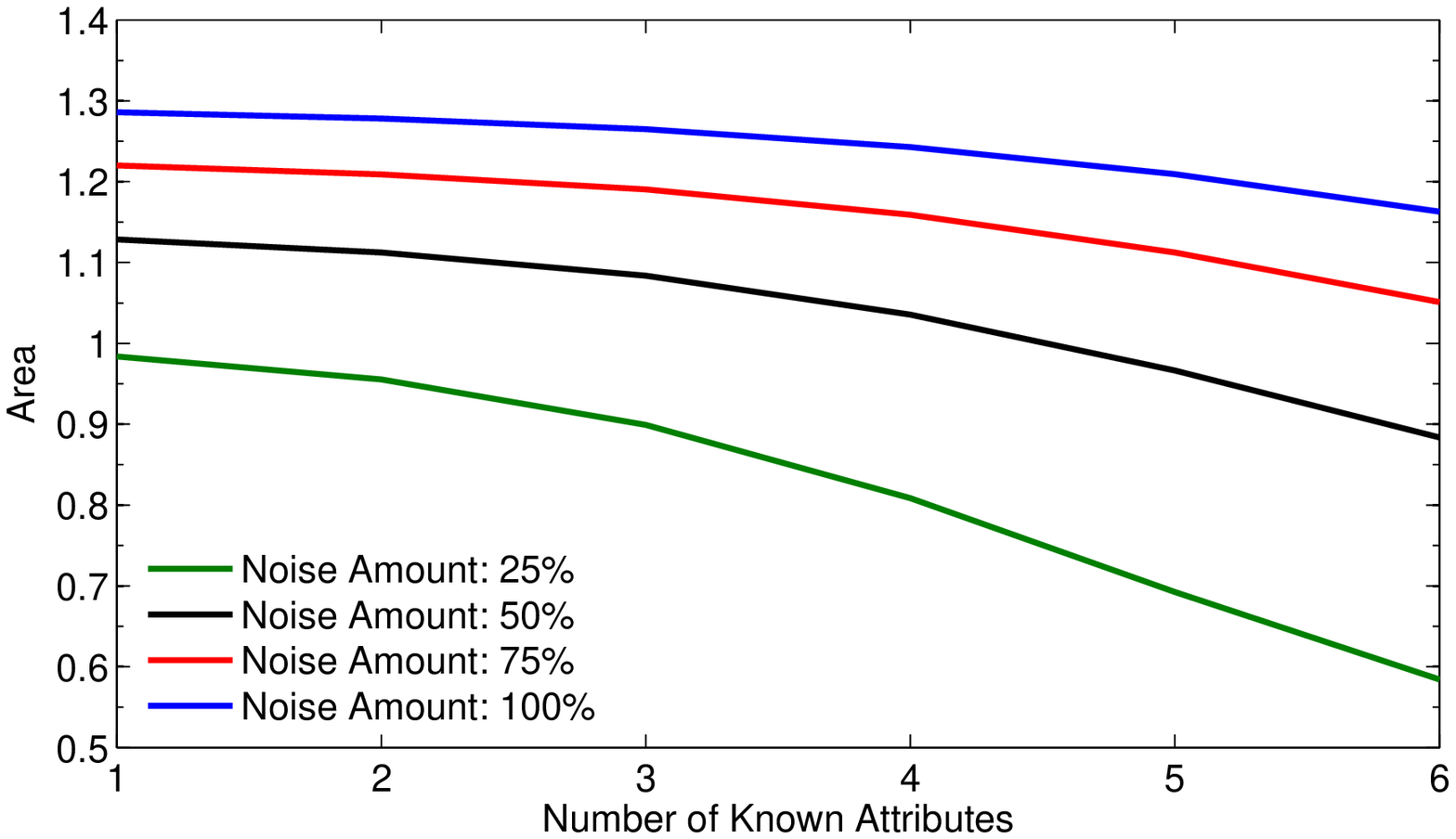}
                    }
    \caption{Sampling, Query Restriction and Noise Addition}
    \label{Sampling_USA_General_H0_Area_vs_AttrNumber}}
\end{figure}

In order to study the effect of sample size on the security we use
four different sampling factors: $5\%,10\%,20\%,50\%$. For each
sample size, we generate $30$ different sample files.
Additionally, we study the effect of the intruder's supplementary
knowledge. We start with supplementary knowledge as little as one
attribute and extended it to reach all attributes except the
confidential one. For each attribute we performed experiments for
all possible values. The results in
Figure~\ref{Sampling_USA_General_H0_Area_vs_AttrNumber} are the
averaged over all 30 samples, all attributes and all values.

\textbf{Query Restriction.} In this experiment, we consider a scenario where an intruder
submits a set of range queries to a DBMS. The intruder performs an
analysis using the answers to the submitted queries as well as the
supplementary knowledge with an aim to infer a confidential
attribute value for the given record, e.g salary in PUMS dataset.
We assume that the intruder has build a system of linear equations
out of the responses to range queries. We use $Q=2l$ to denote the
query set size for the queries a user (i.e., an intruder) is
permitted to ask. For simplicity, we only consider
even query set sizes. We use $k$ to denote the number of queries
and thus also the number of linear equations: $k = \lfloor\frac{2n}{Q}\rfloor-1$.

%

We run the experiment for $5$ different query set size
$\{2,4,8,16,32\}$. For each query set size, we shuffle the records
in the original dataset to get different systems of linear
equations. For each query set size, we produce randomly $30$
different systems of linear equations. Just like in the case of
sampling, for each attribute we performed experiments for all
possible values. The results in Figure~\ref{Sampling_USA_General_H0_Area_vs_AttrNumber} are
the average results, over all $30$ systems of linear equations,
all SK attributes and all values.


 \textbf{Noise Addition.} We consider a scenario where a DBMS alters an original dataset by adding certain level of noise.
 The noise is added to all attributes in the dataset, sensitive and  non-sensitive, categorical and numerical.
 We use additive noise studied in~\cite{Kim:Method86,Tendick:Optimal91,Fuller:Masking93,KimWinkler:Masking95,Yancey:disclos02}.
 The amount of noise is drawn randomly from binomial probability distribution as the nature of attributes in our dataset is discrete.
 The DBMS then releases the perturbed version of the dataset and an intruder obtains a copy of it.  The intruder
 performs an analysis using the released perturbed dataset together with the supplementary knowledge with an aim to infer a confidential
 attribute value, e.g salary in PUMS dataset, corresponding to the individual of concern. We assume there is only one confidential attribute;
 it is straightforward to generalize our experiments to cover more than one confidential attribute.

\section{Discussion and Conclusion}

As expected, for all three SDC techniques our privacy measure, CAE, increases with decrease in utility (Figure 2). In Sampling, utility
is proportional to the sample size, in Query Restriction  utility is inversely proportional to the query size, and in Noise Addition utility is inversely proportional to the amount of noise.

The experiments also show that privacy  declines with additional supplementary knowledge that intruder might have, which is expressed on horizontal axis as the number of known attributes. We note that this decline is sometimes gentle and sometimes sharp, depending on the utility which is in  Figure 2 given as a parameter: for low utility privacy only gently declines with supplementary knowledge, while for higher utility the decline is typically sharp.


Importantly, our experiments demonstrate how we can compare different SDC techniques and select the most suitable one for specific application and requirements. For example, in the absence of supplementary knowledge sampling with size 50\% and query restriction with query set size 8  provide similar level of privacy (blue line in Figure (a) and red line in Figure (c)); however, the privacy sharply drops with  supplementary knowledge increases in sampling, while it remains flat in query restriction. Moreover, Figures (b) and (d) that indicate approximate compromise show a slight superiority of sampling in the absence of supplementary knowledge,  but as supplementary knowledge grows sampling becomes much more vulnerable than query restriction.

In summary, unlike previously proposed privacy measures, our novel information theoretic privacy measure (CAE) has the ability to capture approximate compromise; it   can also be applied to any SDC technique, as long as the probabilities of different confidential values can be estimated. In this paper we considered the most common SDC techniques and showed how CAE can be used to evaluate the privacy they offer, and how this privacy relates to both utility and supplementary knowledge.

\bibliographystyle{amsplain}

\clearpage
\section*{Appendix}
\subsection*{Calculation entropy}
We demonstrate on a small example how  $H(\huge{\varepsilon})$ is calculated. Consider as input a set of values $x_{i}$ of a confidential attribute $X$, where for each $x_{i}$ we have a given probability $p_{i}$, where $p_{i}~\geq~0$ and $\Sigma p_{i}=1$. In order to produce our security measure and compute the area, we need to calculate $H(\huge{\varepsilon})$ for each $\huge{\varepsilon}$. In our example, an intruder learns that a confidential variable $X~\in~[1-10]$~has four possible values with probabilities as follow:

\begin{table}[h]
\begin{center}
\begin{tabular}{| l | l | l | l |}
\hline $x_{1}~=~1$      & $x_{2}~=~3$      & $x_{3}~=~8$      & $x_{4}~=~9$ \\
\hline $p_{1}~=~0.15$   & $p_{2}~=~0.1$    & $p_{3}~=~0.7$    & $p_{4}~=~0.05$  \\
\hline
\end{tabular}
 \label{table:NumericalExample}
\end{center}
\end{table}

\begin{enumerate}

\item We start by calculating initial entropy $H(\huge{\varepsilon}=0)$, as the approximate compromise range $\huge{\varepsilon}$ is initially ``$0$''; this is equivalent to Shannon's entropy for given events and their probabilities: $$H(0)   =  \sum_{i=1}^{4} p(x_{i})\cdot log (\frac{1}{p(x_{i})})=  1.319$$

    \begin{figure}[h]
       \begin{center}
            \includegraphics [scale=0.50]{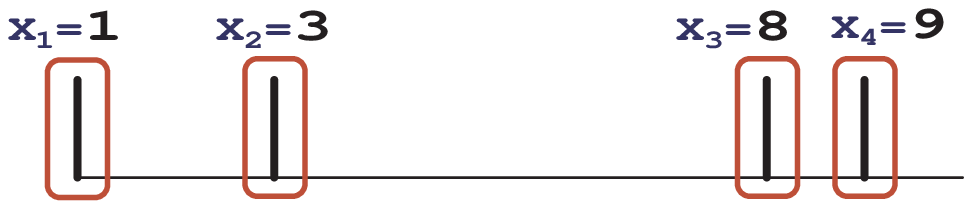}
       \end{center}
    \end{figure}




 \item We calculate the minimum entropy when $\huge{\varepsilon}=1$, that is,~$H(\huge{\varepsilon}=1)$. We cover $x_{3}$ and $x_{4}$ with window of length $1$ and we obtain a combined value $x_{3,4}=8.5$ with probability $p_{3,4}=p_{3}+p_{4}=0.75$:
     $$H (1) = p{1}\cdot log (\frac{1}{p{1}})~+~p{2}\cdot log (\frac{1}{p{2}})~+~(p{3}+~p{4})\cdot log (\frac{1}{(p{3}+~p{4})}) = 1.054$$

    \begin{figure}[h]
       \begin{center}
            \includegraphics [scale=0.50]{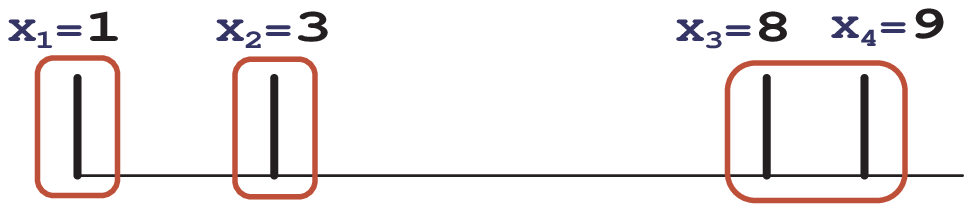}
       \end{center}
    \end{figure}


 \item We calculate the minimum entropy when $\huge{\varepsilon}=2$, that is,~$H(\huge{\varepsilon}=2)$. We cover $x_{3}$ and $x_{4}$ with window of length $1$ and $x_{1}$ and $x_{2}$ with window of length $2$:
$$H(2) = H (X) = (p{1}+~p{2})\cdot log (\frac{1}{(p{1}+~p{2})})~+~(p{3}+~p{4})\cdot log (\frac{1}{(p{3}+~p{4})}) = 0.811$$

    \begin{figure}[h]
       \begin{center}
            \includegraphics [scale=0.50]{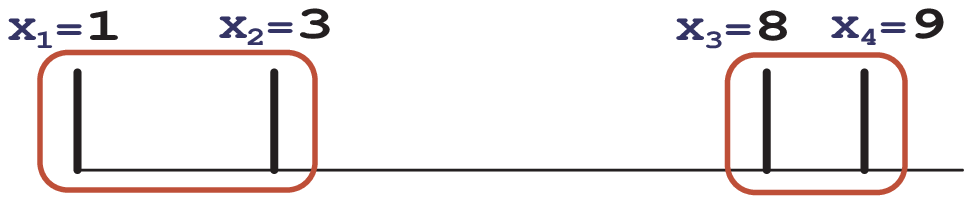}
       \end{center}
    \end{figure}


 \item When $\huge{\varepsilon}=3$ or $4$, we can not do better than for $\huge{\varepsilon}=2$.

 \item We calculate the minimum entropy when $\huge{\varepsilon}=5$, that is,~$H(\huge{\varepsilon}=5)$. Here we have $2$ options so as how to cover the values with window of length $5$ or less. We choose with the option that gives us the minimum entropy, that is, covering $x_{3}$ and $x_{4}$ with window of length $1$ and $x_{1}$ and $x_{2}$ with window of length $2$:
$$H(5) = H (X) = p{1}\cdot log (\frac{1}{p{1}})~+(p{2}+~p{3})\cdot log (\frac{1}{(p{2}+~p{3})})~ +~p{4}\cdot log (\frac{1}{p{4}}) = 0.884$$

    \begin{figure}[h]
       \begin{center}
            \includegraphics [scale=0.50]{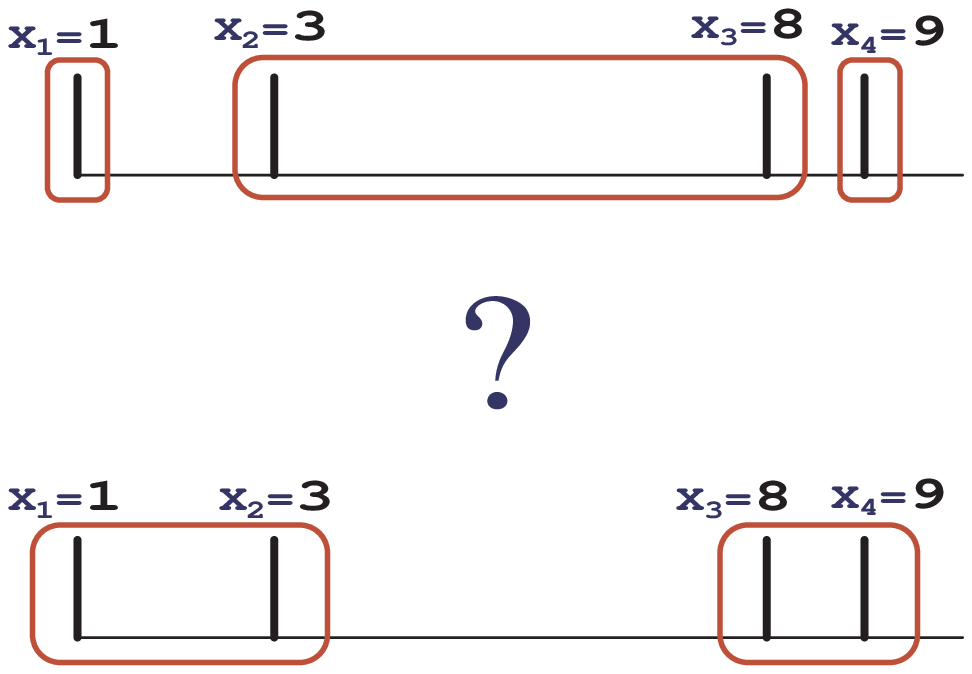}
       \end{center}
    \end{figure}

    $$H(5) = H (X) = (p{1}+~p{2})\cdot log (\frac{1}{(p{1}+~p{2})})~+~(p{3}+~p{4})\cdot log (\frac{1}{(p{3}+~p{4})}) = 0.811$$


 \item We calculate the minimum entropy when $\huge{\varepsilon}=6$, that is,~$H(\huge{\varepsilon}=6)$. We cover $x_{2},x_{3}$ and $x_{4}$ with window of length $6$:
$$H(6) = p{1}\cdot log (\frac{1}{p{1}})~+(p{2}+~p{3}+~p{4})\cdot log (\frac{1}{(p{2}+~p{3}+~p{4})}) = 0.61$$

    \begin{figure}[h]
       \begin{center}
            \includegraphics [scale=0.50]{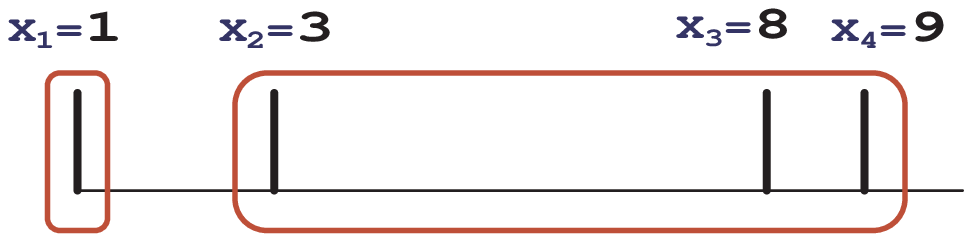}
       \end{center}
    \end{figure}


 \item We calculate the minimum entropy when $\huge{\varepsilon}=7$,~$H(\huge{\varepsilon}=7)$. We cover $x_{1},x_{2}$ and $x_{3}$ with window of length $7$:
$$H(7) = (p{1}+~p{2}+~p{3})\cdot log (\frac{1}{(p{1}+~p{2}+~p{3})})~+~p{4}\cdot log (\frac{1}{p{4}}) = 0.286$$

    \begin{figure}[h]
       \begin{center}
            \includegraphics [scale=0.5]{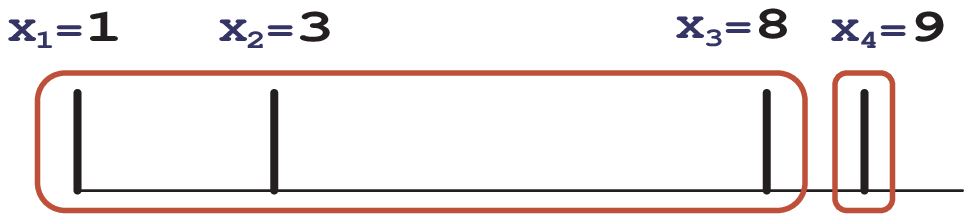}
       \end{center}
    \end{figure}


 \item We calculate the minimum entropy when $\huge{\varepsilon}=8$,~$H(\huge{\varepsilon}=8)$. The window of length $8$ is wide enough to cover all the values and thus will produce an entropy of zero:
$$H(8) = (p{1}+~p{2}+~p{3}+~p{4})\cdot log (\frac{1}{(p{1}+~p{2}+~p{3}+~p{4})}) = 0$$

    \begin{figure}[h]
       \begin{center}
            \includegraphics [scale=0.5]{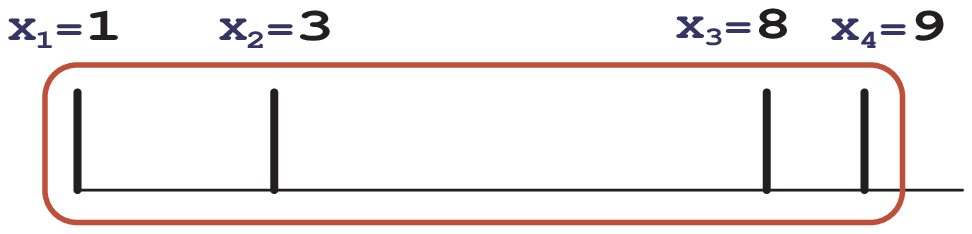}
       \end{center}
    \end{figure}


\end{enumerate}



   \begin{figure}[h]
       \begin{center}
            \includegraphics [scale=0.5]{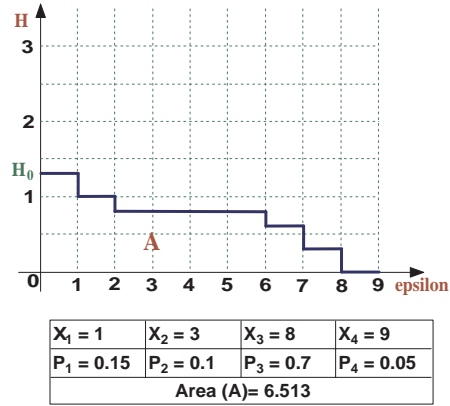}
            \caption{Our security measure for the numerical example.} \label{fig:OurMeasureExampleOutput}
       \end{center}
   \end{figure}


Figure~\ref{fig:OurMeasureExampleOutput}~shows~$H(\huge{\varepsilon})$ and the computed area ($A$) for the above example. This calculation illustrates that computing the minimum entropy $H$ as a function of $\huge{\varepsilon}$ is not a straightforward  task, as we have to make choices. In order to solve this problem efficiently, we design a dynamic programming algorithm to find the optimal entropy for each $\huge{\varepsilon}$ as a measure of disclosure risk (see Section 4).


\subsection*{PUMS Dataset}
Our first data set is the Public Use
Microdata Sample (PUMS)~\cite{USADataset} (experiments on other datasets are ommitted due to length restriction of the paper).
It is a sample of the
actual responses to the American Community Survey (ACS) and is
offered by the U.S. Census Bureau. We have chosen Illinoi state
sample for the year $2006$ as a dataset. The dataset consists of
$2500$ records. Each record is described by $7$ selected
attributes that are shown in Table~\ref{table:PUMS}. The first
column in the table shows the selected attributes. The minimum and
maximum value for each attribute are shown in the second and third
columns. For example, the minimum value for ``$Salary$'' is $10$
while the maximum is $250$. The fourth column indicates the data
type for the corresponding attribute and it can be one of the
following: Categorical, Numerical-Integer, or Numerical-Real. The
next column, Rounding, indicates if the corresponding attribute
values have been rounded. For example, for the attribute
``$Salary$'', we round values to the nearest $10~K$. Symbol "-"
indicates no rounding. The last column contains number of values
in the actual domain for the corresponding attribute after
rounding. We select the first attribute, ``$Salary$'', to be
sensitive (confidential) and hence we need to keep its value
confidential.


\begin{table}[hbtp]
\centering \caption{The selected attributes from PUMS dataset}
\label{table:PUMS}\footnotesize
\begin{center}
\begin{tabular}[]{lccccc}
 \hline
 ~Attribute~~&~~Min.~&~Max.~&~Data Type~&~Rounding~&~No of Values~\\
 \hline

  ~Salary              & $10$   & $250$  & $Numerical-Integer$        & $10$  &  $25$     \\
  ~Age                 & $16$   & $84$   & $Numerical-Integer$        & $-$   &  $69$     \\
  ~Sex                 & $1$    & $2$    & $Categorical$              & $-$   &  $2$      \\
  ~Education           & $1$    & $16$   & $Categorical$              & $-$   &  $16$     \\
  ~Industry            & $1$    & $18$   & $Categorical$              & $-$   &  $18$     \\
  ~Occupation          & $1$    & $25$   & $Categorical$              & $-$   &  $25$     \\
  ~Work Travel Time    & $1$    & $177$  & $Numerical-Integer$        & $-$   &  $177$    \\

 \hline
\end{tabular}
\end{center}
\end{table}

\subsection*{Details of Experiments}

We use the following notation: (1) $n$~is a number of records in the original database;
(2) $C$~is the confidential attribute;
(3) $c_i$, $1 \leq i \leq n$,  is the value of the confidential attribute $C$ in record $i$;
(4) $D$~is the domain of the confidential attribute $C$ in the original database;
(5) $d_{i}$~is the $i^{th}$ value of the confidential attribute in the domain $D$, where $1\leq i \leq |D|$;
(6) $p_{i}$~is the probability that the confidential attribute value is $d_{i}$;
(7) $M_{o}$~is the set of records in the original dataset that matches the intruder supplementary
  knowledge.
	
\textbf{Sampling.} We consider a scenario where an intruder obtains a copy of the
released sample file and performs an analysis using the
released sample together with the supplementary knowledge with an
aim to infer a confidential attribute value corresponding to the
individual in question, e.g., salary in PUMS dataset.

We introduce the following notation specific to sampling:
(1)  $M_{s}$~is the set of records in the released sample dataset
  that matches the intruder's supplementary knowledge
(2) $f_{d_{i}}$~is the frequency of $d_{i}$ in $M_{s}$, that is,
  how many times $d_{i}$ appears in records that belong to $M_{s}$.

To compute the initial and average entropy, we need a set
of $d_{i}$'s and their corresponding $p_{i}$'s. We first identify
$|M_{s}|$ and $|M_{o}|$ and then calculate $f_{d_{i}}$ for each
$d_i$. Then for each $d_i$ we compute a corresponding $p_{i}$,
that is, the probability that the record in question is in the
sample and $d_{i}$ is its confidential value $or$ that the record
does not appear in the sample and $d_{i}$ is its confidential
value. Note that for those records that do not appear in the sample the
equal probability of all $|D|$ values is assumed. We also assume
that $|M_{o}|$ is a part of the intruder's supplementary
knowledge.
\begin{eqnarray}
                    \nonumber  p_{i} = \frac{f_{d_{i}}}{|M_{s}|} \cdot \frac{|M_{s}|}{|M_{o}|} + \frac{|M_{o}|-|M_{s}|}{|M_{o}|} \cdot|D|
\end{eqnarray}

\textbf{Query Restriction.} The intruder obtains a system of $k$ linearly independent
equations in $n$ unknowns. To be able to solve the system and
completely compromise the database, the intruder needs $n$
linearly independent equations. Nevertheless, with $k < n$
linearly independent equations, the intruder is able to find the
upper and lower bounds (min, max) for the confidential attribute
value in each record.

We follow the evaluation of an entropy based measure of disclosure
risk presented in~\cite{Oganian-Ferrer:Posteriori03} and solve two
linear programming problems, maximisation and minimisation, to
find $L$ and $U$, the upper and the lower bound for $r_i$, the
value of the confidential attribute $C$ in record $i$ that matches
the intruder's SK. The constraints for the linear programming
problems consist of the given system of $k$ linearly independent
equations in $n$ unknowns, plus a system of inequalities of the
form $r_i \geq d_{min}$ and $r_i \leq d_{max}$, where $d_{min}$
and $d_{max}$ are the minimum and the maximum value in the domain
$D$. The linear function to be maximised (minimised) is the
confidential value in the record $i$. We use $L$ and $U$ for each
record that matches the intruder's SK to find the probability
$p_{i}$ for each value $d_i$ in the domain $D$. See
Algorithm~\ref{algorithm:QR} bellow.

\textbf{Noise Addition.} Algorithm 3  shows the steps
that are followed in order to analyse noise and obtain a set of
$d_{i}$ and their corresponding $p_{i}$. Figure 4 shows the average  of initial and average entropy over $30$ perturbed files,  each size of the intruder's SK, and for each amount of noise. By $k\%$ noise, we mean that the maximum noise $M$ is $k\%$ of $|D|-1$, rounded to the nearest $2$.

\begin{algorithm}[H] \footnotesize{
  \SetLine
  \KwOut{$d[~],p[~]$}
  \tcp{Find the set of records in the original dataset that matches the intruder's SK.}
   $Find~M_{o}$\;
  \ForEach{$i$~in~$[1,|D|]$}{
        $p_{i} \leftarrow 0$\;
  }
  \ForEach{record belonging to $M_{o}$}{
            \tcp{Find $L$ and $U$, that is, the lower and the upper bound for the confidential attribute value in the current record.}
            $Find~L~and~U$\;

            \ForEach{$i$~such~that~$d_i$~in~$[L,U]$}{
                $p_{i} \leftarrow p_{i} + (\frac{1}{U-L+1} \cdot \frac{1}{|M_{o}|})$\;
            }

  }

  \textbf{Display:} $d[~],~p[~]$
  \caption{Finding $d_{i}$ and their corresponding $p_{i}$}
 \label{algorithm:QR}}
\end{algorithm}

\begin{algorithm}[H] \footnotesize{
  \SetLine
  \KwOut{$d[~],p[~]$}
  \tcp{Find the set of records in the original dataset that matches the intruder's SK.}
   $Find~M_{o}$\;
  \ForEach{$i$~in~$[1,|D|]$}{
        $p_{i} \leftarrow 0$\;
  }
  \ForEach{record belonging to $M_{o}$}{
            \tcp{Find $L$ and $U$, that is, the lower and the upper bound for the confidential attribute value in the current record.}
            $Find~L~and~U$\;

            \ForEach{$i$~such~that~$d_i$~in~$[L,U]$}{
                $p_{i} \leftarrow p_{i} + (\frac{1}{U-L+1} \cdot \frac{1}{|M_{o}|})$\;
            }

  }

  \textbf{Display:} $d[~],~p[~]$
  \caption{Finding $d_{i}$ and their corresponding $p_{i}$}
 \label{algorithm:QR}}
\end{algorithm}

\begin{figure}[H]
    \raggedleft
        \subfloat[$H_{0}$ vs Intruder's SK ]{%
                    \label{fig:NA_USA_General_H0_vs_AttrNumber}%
                    \includegraphics[width=0.5\textwidth]{General_H0_vs_AttrNumber_30Files_NA.eps}  }%
        \subfloat[Area vs Intruder's SK ]{%
                    \label{fig:NA_USA_General_Area_vs_AttrNumber}%
                    \includegraphics[width=0.5\textwidth]{General_Area_vs_AttrNumber_30Files_NA.eps}
                    }\\
        \subfloat[$H_{0}$ vs Noise Amount]{%
                    \label{fig:NA_USA_General_H0_vs_NoiseAmnt}%
                    \includegraphics[width=0.5\textwidth]{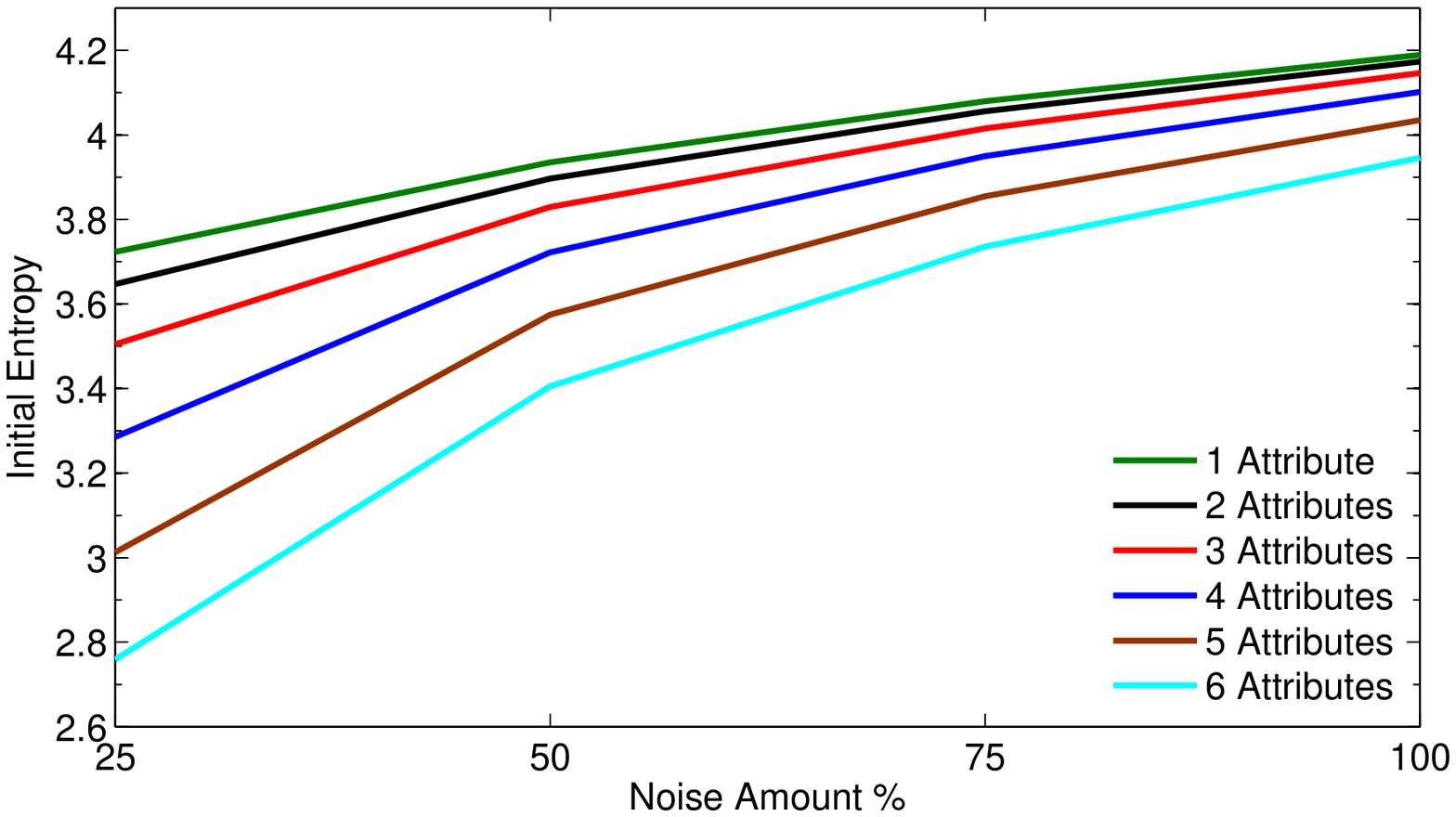}  }%
        \subfloat[Area vs Noise Amount]{%
                    \label{fig:NA_USA_General_Area_vs_NoiseAmnt}%
                    \includegraphics[width=0.5\textwidth]{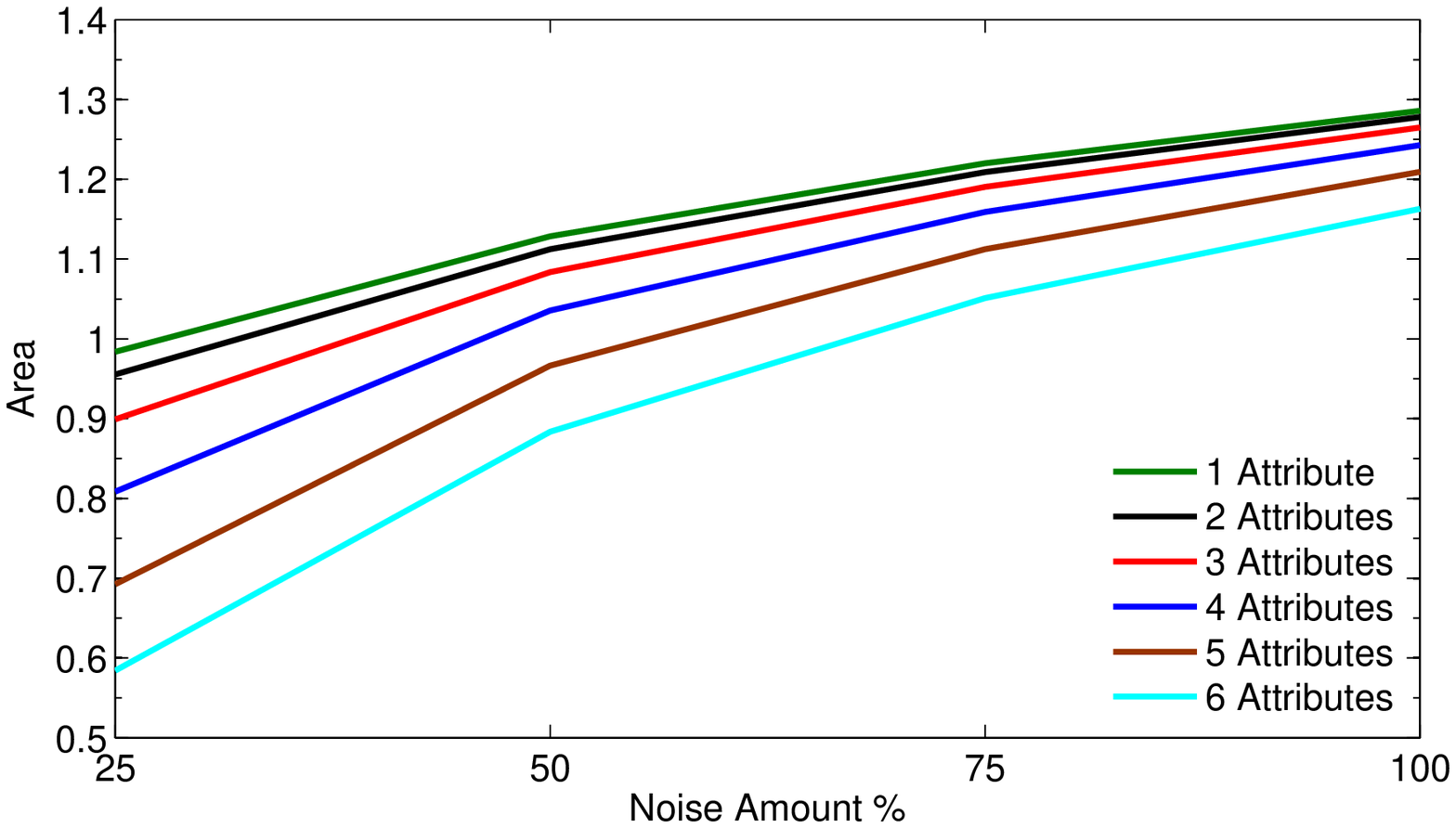}
                    }\\
    \caption{SDC: Noise Addition, Dataset: PUMS.}
    \label{NA_USA_General}
\end{figure}

\end{document}